\documentclass[12pt,a4paper]{article}
\usepackage{fix-cm}
\usepackage[a4paper,left=2.0cm,right=2.0cm, bottom=2.5cm]{geometry}
\usepackage{ifthen} 
\usepackage{subfig}
\usepackage{graphicx}
\usepackage{longtable}
\usepackage[abs]{overpic}
\usepackage{multirow}
\usepackage{braket}
\usepackage{lipsum}
\usepackage{comment}
\usepackage{amsmath}
\usepackage{amssymb}
\usepackage{xspace}
\usepackage{orcidlink}
\usepackage{pgfgantt}
\usepackage{siunitx}
\usepackage{multicol}
\DeclareSIUnit\clight{\text{\ensuremath{c}}}
\DeclareSIUnit\electronvolt{e\kern-0.15ex V} 

\usepackage{booktabs}
\usepackage{lineno}

\usepackage{upgreek}
\usepackage[title]{appendix}

\usepackage{color}   
\usepackage{hyperref}
\hypersetup{linktocpage}
\hypersetup{
    colorlinks=true, 
    citecolor=blue,
    linkcolor=blue,     
}

\newboolean{pdflatex}

\newboolean{uprightparticles}
\setboolean{uprightparticles}{false} 

\newcommand*{\GeVc} {\ensuremath{\text{Ge\kern -0.1em V/}c} }

\cleardoublepage

\newcommand{\appendixnumberline}[1]{Appendix\space}

\let\oldappendix\appendix
\makeatletter
\renewcommand{\appendix}{%
  \addtocontents{toc}{\let\protect\numberline\protect\appendixnumberline}%
  \renewcommand{\@seccntformat}[1]{Appendix~\csname the##1\endcsname\quad}%
  \oldappendix
}
\makeatother

\begin{document}

\begin{titlepage}
\begin{figure}
    \includegraphics[width=0.1\textwidth]{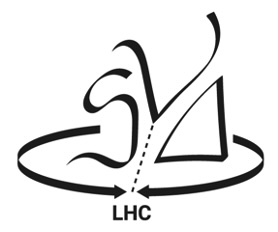}
\end{figure}
\vspace*{-3.0cm}

\hspace*{-0.5cm}
\begin{tabular*}{\linewidth}{lc@{\extracolsep{\fill}}r}
\end{tabular*}


{\bf \Large
\begin{center}
{\bf\huge Input from the SND@LHC collaboration to the 2026 Update 
 to the European Strategy for Particle Physics 
}
\end{center}
}

\vspace*{1.0cm}

\begin{center}
SND@LHC Collaboration
\end{center}

\begin{center}
{\bf \today }\\
\end{center}

By observing collider neutrino interactions of different flavours, the SND@LHC and Faser$\nu$ experiments have shown that the LHC can make interesting contributions to neutrino physics. This document summarizes why the SND@LHC Collaboration intends to continue taking data at the High Luminosity LHC (HL-LHC).

The upgraded detector\cite{SND@HL-LHC_TP} will  instrument the regions of both the neutrino vertex and the magnetized calorimeter with silicon microstrips. The use of this technology will allow us to continue the physics program of the current SND@LHC detector with higher statistics. It will also offer new possibilities. For instance, the magnetization of the hadron calorimeter will enable the separation between neutrinos and antineutrinos. This could lead to the first direct observation of tau antineutrinos.

The use of ultrafast timing layers will enable triggers to be sent to ATLAS, potentially allowing the identification of the charm quark pair that produced the neutrino interacting in the detector. Such tagging of the neutrino source would fulfill Pontecorvo's original proposal of a tagged neutrino beam.
The experiment will perform unique measurements with high energy neutrinos and will also provide a means to measure gluon parton distribution functions in a previously unexplored domain  (Bjorken-$x < 10^{-5}$).

Furthermore, the technological advancements of the upgrade and the experience that will be gained in the areas of operation and data analysis will play a crucial role in the design of the neutrino detector for the SHiP experiment.

\end{titlepage}

\cleardoublepage



%
\vspace*{1mm}
\begin{flushleft}
\begin{center}
D.~Abbaneo$^{9}$\orcidlink{0000-0001-9416-1742},
C.~Ahdida$^{9}$,
S.~Ahmad$^{42}$\orcidlink{0000-0001-8236-6134},
R.~Albanese$^{1,2}$\orcidlink{0000-0003-4586-8068},
A.~Alexandrov$^{1}$\orcidlink{0000-0002-1813-1485},
F.~Alicante$^{1,2}$\orcidlink{0009-0003-3240-830X},
F.~Aloschi$^{1,2}$,
N.~Amapane$^{14,15}$\orcidlink{0000-0001-9449-2509},
M.~Andreini$^{9}$,
K.~Androsov$^{6}$\orcidlink{0000-0003-2694-6542}
% ...
\end{center}
\vspace{0.5cm}
\begin{footnotesize}
\textit{
$^{1}$Sezione INFN di Napoli, Napoli, 80126, Italy\linebreak
$^{2}$Universit\`{a} di Napoli ``Federico II'', Napoli, 80126, Italy\linebreak
$^{3}$Affiliated with an institute formerly covered by a cooperation agreement with CERN\linebreak
$^{4}$Sezione INFN di Bologna, Bologna, 40127, Italy\linebreak
$^{5}$Universit\`{a} di Bologna, Bologna, 40127, Italy\linebreak
$^{6}$Institute of Physics, EPFL, Lausanne, 1015, Switzerland\linebreak
$^{7}$Physik-Institut, UZH, Z\"{u}rich, 8057, Switzerland\linebreak
$^{8}$Hamburg University, Hamburg, 22761, Germany\linebreak
$^{9}$European Organization for Nuclear Research (CERN), Geneva, 1211, Switzerland\linebreak
$^{10}$Laboratory of Instrumentation and Experimental Particle Physics (LIP), Lisbon, 1649-003, Portugal\linebreak
$^{11}$Faculty of Physics,Sofia University, Sofia, 1164, Bulgaria\linebreak
$^{12}$Universit\`{a} degli Studi di Cagliari, Cagliari, 09124, Italy\linebreak
$^{13}$University of Leiden, Leiden, 2300RA, The Netherlands\linebreak
$^{14}$INFN, Sezione di Torino, Torino, Italy\linebreak
$^{15}$Universit\`{a} di Torino, Torino, Italy\linebreak}
$^{16}$Universit\`{a} di Napoli Parthenope, Napoli, 80143, Italy\linebreak
$^{17}$Sungkyunkwan University, Suwon-si, 16419, Korea\linebreak
$^{18}$Institut f\"{u}r Physik and PRISMA Cluster of Excellence, Mainz, 55099, Germany\linebreak
$^{19}$Humboldt-Universit\"{a}t zu Berlin, Berlin, 12489, Germany\linebreak
$^{20}$Universit\`{a} del Sannio, Benevento, 82100, Italy\linebreak
$^{23}$Middle East Technical University (METU), Ankara, 06800, Turkey\linebreak
$^{24}$Universit\`{a} della Basilicata, Potenza, 85100, Italy\linebreak
$^{26}$Imperial College London, London, SW72AZ, United Kingdom\linebreak
$^{27}$Millennium Institute for Subatomic physics at high energy frontier-SAPHIR, Santiago, 7591538, Chile\linebreak
$^{29}$Department of Physics Education and RINS, Gyeongsang National University, Jinju, 52828, Korea\linebreak
$^{30}$Gwangju National University of Education, Gwangju, 61204, Korea\linebreak
$^{31}$Nagoya University, Nagoya, 464-8602, Japan\linebreak
$^{32}$Center for Theoretical and Experimental Particle Physics, Facultad de Ciencias Exactas, Universidad Andr\`es Bello, Fernandez Concha 700, Santiago, Chile\linebreak
$^{33}$Korea University, Seoul, 02841, Korea\linebreak
$^{34}$Toho University, Chiba, 274-8510, Japan\linebreak
$^{35}$Niels Bohr Institute, Copenhagen, 2100, Denmark\linebreak
$^{36}$Present address: Faculty of Engineering, Kanagawa, 221-0802, Japan\linebreak
$^{37}$Constructor University, Bremen, 28759, Germany\linebreak
$^{38}$Chulalongkorn University, Bangkok, 10330, Thailand\linebreak
$^{39}$Chiang Mai University , Chiang Mai, 50200, Thailand\linebreak
$^{40}$Departamento de F\'isica, Facultad de Ciencias, Universidad de La Serena, La Serena, 1200, Chile \linebreak
$^{41}$Also at: Universit\`{a} di Pisa, Pisa,  56126, Italy \linebreak
$^{42}$Present address: Pakistan Institute of Nuclear Science and Technology (PINSTECH), Nilore, 45650, Islamabad, Pakistan \linebreak
$^{43}$Also at: Atilim University, Ankara, Turkey\linebreak
$^{44}$Present address: Pegaso University, Napoli, Italy\linebreak
$^{45}$Laboratori Nazionali dell'INFN di Gran Sasso, L'Aquila, Italy\linebreak
$^{46}$Universit\`{a} di Salerno, Salerno, 84084, Italy\linebreak
$^{47}$INFN, Sezione di Pisa, I-56127 Pisa, Italy\linebreak
$^{48}$INFN, Sezione di Padova, Padova, Italy\linebreak
$^{49}$Universit\`{a} di Padova, Padova, Italy\linebreak
$^{50}$Universit\`{a} degli Studi di Milano-Bicocca, Milano Italy\linebreak
$^{51}$INFN, Sezione di Milano-Bicocca, Milano, Italy\linebreak
$^{52}$Physics Department, Boston University, USA\linebreak
\end{footnotesize}
\end{flushleft}

\clearpage

\thispagestyle{empty}

\begingroup\baselineskip.99\baselineskip
\setcounter{tocdepth}{3}
\endgroup
\clearpage
\newpage 

\pagenumbering{arabic}
\setcounter{page}{1}
\section{Introduction }
\label{sec:introduction}

\begin{figure}[h!]
    \centering
    \includegraphics[width=1.0\linewidth]{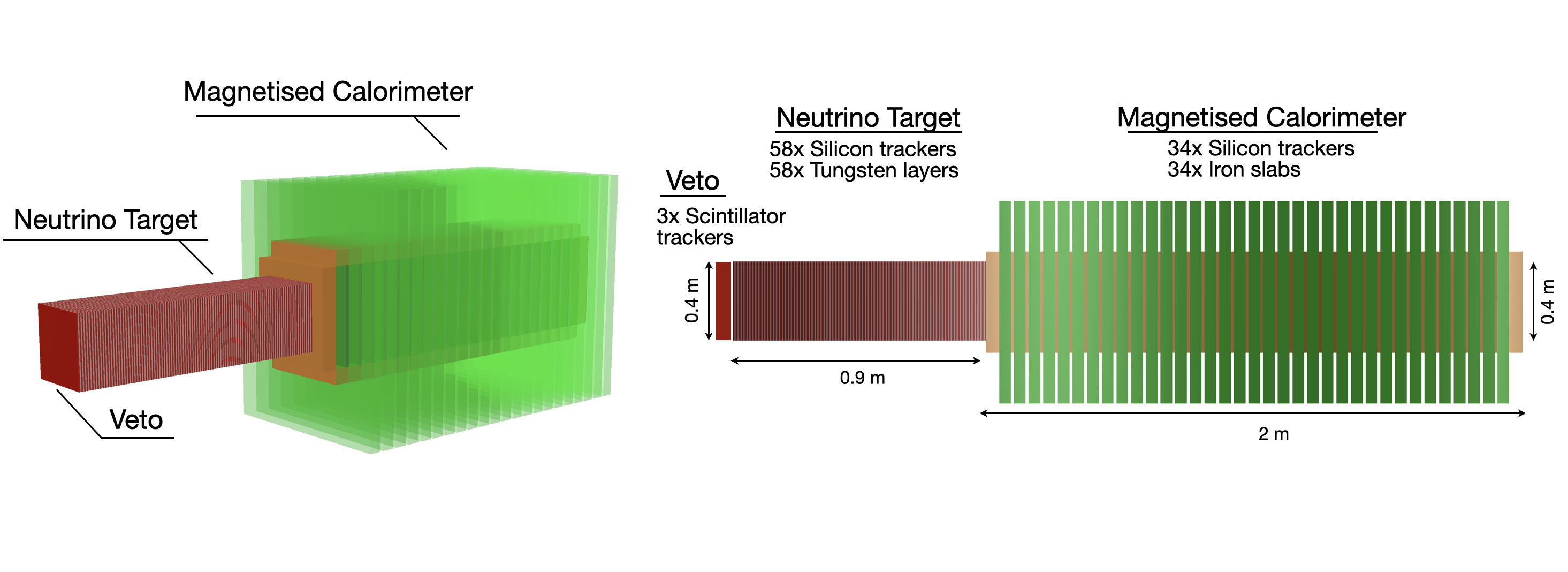}
    
    \vspace{-1cm}
    \caption{Layout of the upgraded SND@LHC detector, as implemented in the 3D simulation.}
    \label{fig:detector_layout}
\end{figure}

The upgraded SND@LHC detector for the HL-LHC phase (hereafter named SND@HL-LHC for brevity) is designed to identify all three neutrino flavors and to measure their energy. It consists of a target and a magnetized calorimeter (Figure~\ref{fig:detector_layout}). The Tungsten-Silicon sandwich of the target serves as both an electromagnetic calorimeter and a vertex detector, distinguishing neutrino interaction and tau decay vertices. The Iron-Silicon sandwich of the magnetized calorimeter measures hadronic energy, identifies muons from $\nu_\mu$ interactions and tau decays, and determines their energy.

Compared to the current SND@LHC detector, the proposed setup features three major upgrades: 
\begin{enumerate}
\item The use of silicon strip modules recovered from the decommissioned CMS Outer Barrel tracker to equip both  the  vertex detector and the magnetic spectrometer as active element. 
\item The addition of four fast timing detector layers in the target  to trigger the readout of the silicon detectors and  to  send a trigger to ATLAS for the tagging of the neutrino parent. 
\item The magnetized hadron calorimeter will allow for precise ($\approx 20\% $ at 1 TeV) muon momentum measurements.
\end{enumerate}

The existing veto system will be used to reject charged particles and tag muon for calibration purposes.

The target consists of 58 tungsten plates (7 mm thick) alternating with 58 silicon tracker layers, covering $40$x$40$ cm$^2$, totaling 1.3 tons and 116 radiation lengths ($X_0$). The segmented electromagnetic calorimeter ensures efficient  $\nu_e$ CC interaction identification.

The magnetized hadron calorimeter has 34 iron walls (5 cm thick) interleaved with silicon tracker layers, totaling 8 interaction lengths ($\lambda_{int}$). A 1.8~T vertical magnetic field enables muon charge and momentum measurement, distinguishing neutrinos from anti-neutrinos and enhancing physics capabilities.

The proposed detector is designed to fit entirely within the existing TI18 tunnel. 

It is worth noting that a modest civil engineering work, with an excavation of 4.5 cubic meters, would  improve by a factor $\approx7$ the accumulated statistics and hence the physics reach, in particular in the  $\nu_\tau$ sector.

\section{Physics goals}
\label{sec:overview}

\begin{figure}[htbp]
\centering
\includegraphics[width=0.8\columnwidth]{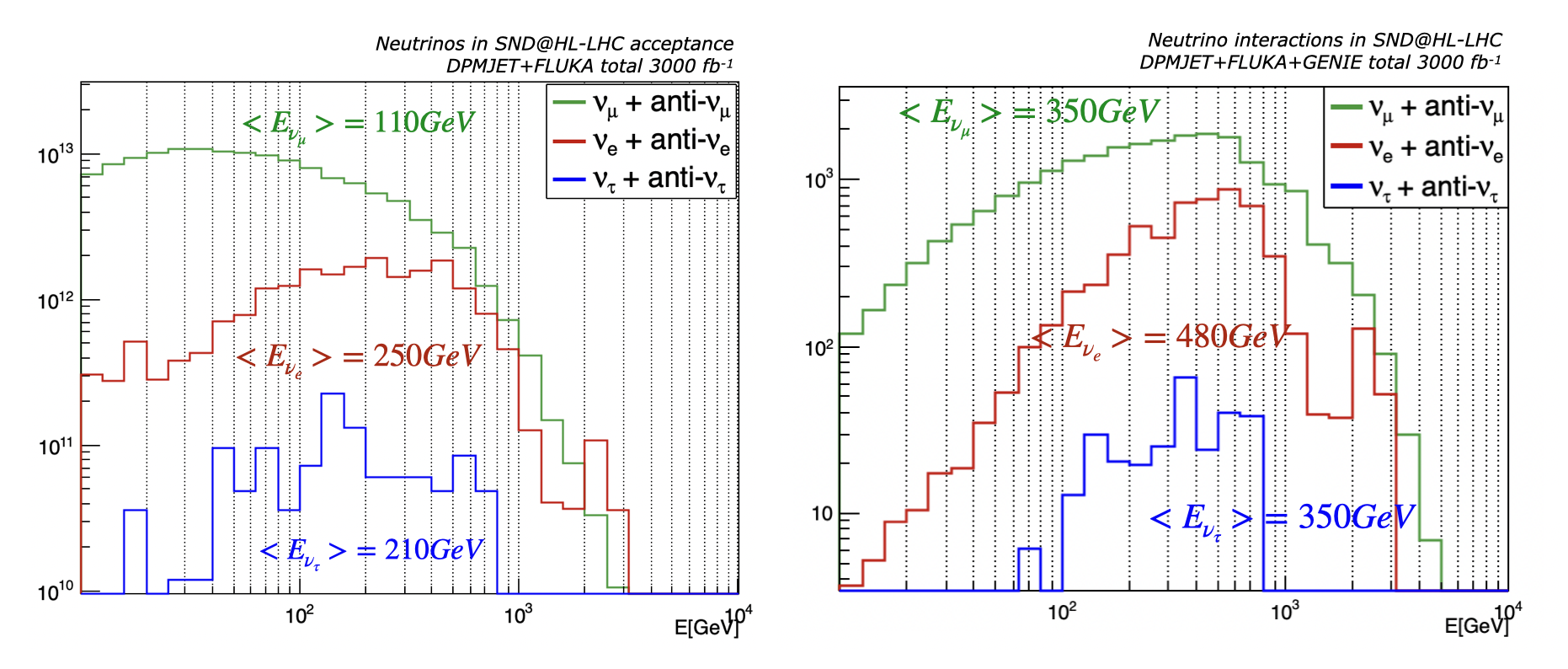}
\caption{Energy spectra of the three neutrino flavours in the target acceptance (left) and undergoing CC~DIS interactions in the target (right). The normalization corresponds to 3000 fb$^{-1}$. Average energies are also reported.}
\label{fig:nu_spectra}
\end{figure}

SND@HL-LHC aims to exploit the large neutrino flux in the forward region of $pp$ collisions at HL-LHC to study neutrino production and interaction mechanisms. Its large cross-sectional area covers a broad pseudo-rapidity range, enabling measurements at lower pseudo-rapidities where charm hadron decays dominate. The experiment, over the full HL-LHC period,  will detect thousands of  $\nu_\mu$ and  $\nu_e$ interactions and a few hundred  $\nu_\tau$ interactions, with energies up to a few TeV. Neutrino spectra are shown in Figure~\ref{fig:nu_spectra}, with CC~DIS interaction spectra in the right panel.

SND@HL-LHC expects a high event rate of all three neutrino flavors, enabling Lepton Flavour Universality (LFU) tests via interaction ratios. The $\nu_\tau$ to $\nu_e$ ratio, dominated by charmed hadron decays, will be statistically limited, while the $\nu_e$ to $\nu_\mu$ ratio will have high precision but systematic uncertainties from pion-decay contamination. \textbf{These complementary measurements will allow precise LFU tests with all three neutrino flavors.}


High-energy  $\nu_e$ from charm decays enable studies of forward charm production in $pp$ collisions, \textbf{probing the gluon Parton Density Function (PDF) at Bjorken $x$ below 10$^{-5}$.} Constraining the gluon PDF in this regime improves cosmic neutrino background models and predictions for future high-energy hadron colliders. Aside from future hadron colliders and HL-LHC neutrino experiments, no other current or planned experiments access such small $x$ values.


High-statistics cross-section measurements will be conducted for the highest-energy neutrinos ever produced by human-made sources. Tau neutrino interactions are particularly significant, with potential for unprecedented event collection. Notably, SND@HL-LHC may \textbf{achieve the first direct observation of the tau antineutrino.} The large  $\nu_\tau$ flux will also \textbf{enable competitive limits on its magnetic moment} via  neutrino elastic scattering on electrons~\cite{DONUT:2001zvi}.

Correlations between neutrinos detected in SND@HL-LHC and ATLAS data will be explored. In particular, a few hundred open charm production events are expected to emit a charm hadron in the ATLAS detector acceptance in coincidence with a neutrino interaction in SND@HL-LHC. \textbf{This charm-tagging approach allows for the selection of a sample of neutrinos with charm-hadron origin.} Such a sample can be used in LFU tests with clean ratios greatly reducing the systematic uncertainty.

Finally, SND@HL-LHC is also sensitive to new physics scenarios, particularly in \textbf{scattering signatures of feebly interacting particles.} This approach relies on searching for excesses in the rates of processes that are well predicted by the Standard Model, such as the ratio of neutral current to charged current neutrino interactions, and neutrino scattering on electrons~\cite{Boyarsky:2021moj}.

\section{Physics performances}
\label{sec:physics}
\subsection{Neutrino yields}
The neutrino yields at the target region for the three different neutrino flavours are reported in the left column of 
Table~\ref{tab:nu_flux}. An integrated luminosity of 3000 fb$^{-1}$ and  the +250 \textmu rad-H beam configuration are assumed.

The expected number of CC and NC neutrino interactions occurring in the detector target assuming a 1.3 ton tungsten mass is reported in  Table~\ref{tab:nu_flux}.

The neutrino component produced in charmed hadron decays was also estimated using the \textsc{Pythia8}~\cite{Pythia8} and \textsc{POWHEG}~\cite{Buonocore:2023kna} generators activating hard QCD processes only, which provide a number of expected neutrino interactions $20\div30\%$ less with respect to \textsc{DPMJET} in the pseudorapidity range of SND@HL-LHC.
Beyond the differences in the expected number of interactions, subtle variations can be observed in the ratio of neutrino to antineutrino fluxes, likely arising from different hadronization models. The ability of SND@HL-LHC to distinguish between $\nu$ and $\bar{\nu}$ will aid in refining these models.

\begin{table}[hbtp]
\centering
\begin{tabular}{c | c  c | c | c  c | c}
\toprule
       & \multicolumn{3}{c|}{CC~DIS} &              \multicolumn{3}{c}{NC~DIS} \\
       & \multicolumn{2}{c|}{\textsc{DPMJET}} & \textsc{POWHEG+P8} & \multicolumn{2}{c|}{\textsc{DPMJET}} & \textsc{POWHEG+P8}\\
Flavour & $\pi/K$ & charm & charm & $\pi/K$ & charm & charm \\
\midrule
$\nu_\mu$        & 9.6$\times 10^3$ &  1.4$\times 10^3$ & 1.5$\times 10^3$ & 3.0$\times 10^3$ & 4.2$\times 10^2$ & 4.4$\times 10^2$\\
$\bar{\nu}_\mu$  & 3.1$\times 10^3$ &  9.9$\times 10^2$ & 6.8$\times 10^2$ & 1.1$\times 10^3$ & 3.7$\times 10^2$ & 2.5$\times 10^2$\\
$\nu_e$          & 5.0$\times 10^2$ &  1.6$\times 10^3$ & 1.5$\times 10^3$ & 1.4$\times 10^2$ & 4.9$\times 10^2$ & 4.5$\times 10^2$\\
$\bar{\nu}_e$    & 2.0$\times 10^2$ &  1.1$\times 10^3$ & 6.9$\times 10^2$ & 6.0$\times 10^1$ & 4.2$\times 10^2$ & 2.5$\times 10^2$\\
$\nu_\tau$       & ---              &  2.2$\times 10^2$ & 9.3$\times 10^1$ & ---              & 7.6$\times 10^1$ & 3.0$\times 10^1$\\
$\bar{\nu}_\tau$ & ---              &  5.7$\times 10^1$ & 5.0$\times 10^1$ & ---              & 2.5$\times 10^1$ & 2.0$\times 10^1$\\
\midrule
 Tot             & 1.3$\times 10^4$ &  5.4$\times 10^3$ & 4.5$\times 10^3$ & 4.3$\times 10^3$ & 1.8$\times 10^3$ & 1.4$\times 10^3$\\
\bottomrule
 \end{tabular}
  \caption{Number of CC~DIS and NC~DIS neutrino interactions, assuming 3000\,fb$^{-1}$, as estimated with \textsc{DPMJET}/\textsc{FLUKA}, \textsc{POWHEG+P8} and \textsc{GENIE} generators.}
  \label{tab:nu_flux}
 \end{table}

It has to be noted that a large number of neutrino interactions will occur also in the HCAL. The high granularity of the silicon trackers will make the upstream part of the HCAL a good target to identify muon- and  $\nu_e$s interactions thus adding to the overall neutrino interactions statistics. 

\begin{figure}[htbp]
\centering
\includegraphics[width=.98\columnwidth]{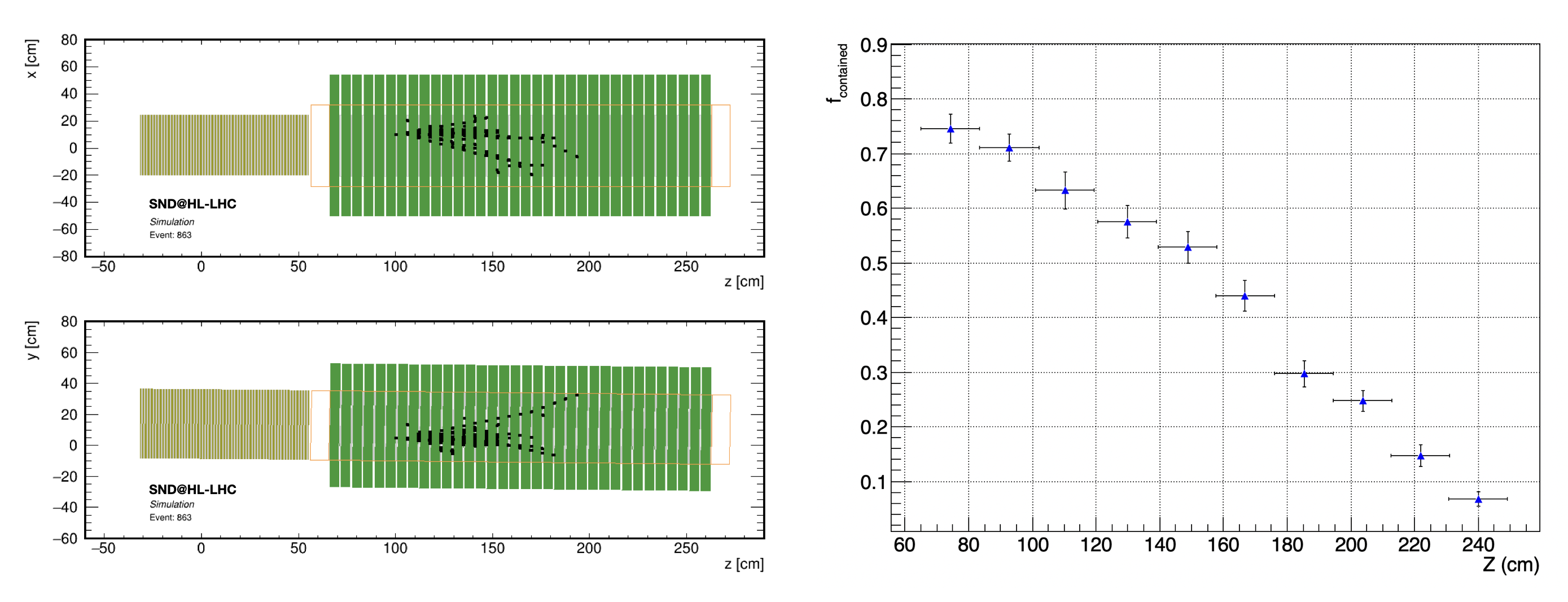}
\caption{Left: display of an  $\nu_e$ interaction occurring in the HCAL. Right: fraction of $\nu_e$ CC~DIS interactions occurring in the HCAL and fully contained in the detector, as a function of the longitudinal position of the interaction.}
\label{fig:containment}
\end{figure}

Preliminary estimates based on the longitudinal development of hadronic and electromagnetic showers in the HCAL show that about 40\% of muon and  $\nu_e$s interacting in Iron layers are fully contained in the detector --- with the exception of the muon produced in $\nu_\mu$ CC~DIS interactions. 
The right panel of Figure~\ref{fig:containment} shows the fraction of $\nu_e$ CC~DIS interactions occurring in the HCAL that have a longitudinal development fully contained in the detector, as a function of the interaction position. A display of one of those is shown in the left panel of the same figure. The corresponding neutrino yield is reported in Table~\ref{tab:containment}, assuming 3000 fb$^{-1}$ and the +250 \textmu rad-H beam configuration. This indicates a potential  increase of about 50\% of detectable neutrino interactions.

 \begin{table}[hbtp]
\centering
\begin{tabular}{c | c | c | c}
\toprule
 Flavour      & Target & HCAL &  Target+HCAL\\
 \midrule
$\nu_\mu$+$\bar{\nu}_\mu$   & 1.5$\times 10^4$ & 8.8$\times 10^3$ & 2.4$\times 10^4$\\
$\nu_e$+$\bar{\nu}_e$       & 3.4$\times 10^3$ & 2.1$\times 10^3$ & 5.5$\times 10^3$\\
$\nu_\tau$+$\bar{\nu}_\tau$ & 2.8$\times 10^2$ & 1.7$\times 10^2$ & 4.5$\times 10^2$\\
\midrule
 Tot                        & 1.9$\times 10^4$ & 1.1$\times 10^4$ & 3.0$\times 10^4$ \\
\bottomrule
 \end{tabular}
  \caption{Number of neutrinos CC~DIS interactions in the Target and in the HCAL, assuming 3000\,fb$^{-1}$ and the +250 \textmu rad-H configuration. Longitudinal shower containment is required for interactions in the HCAL.}
  \label{tab:containment}
 \end{table}

\subsection{Neutrino physics }
\begin{figure}[h!]
    \centering
	\includegraphics[trim={20 20 20 20}, clip, width=0.6\textwidth]{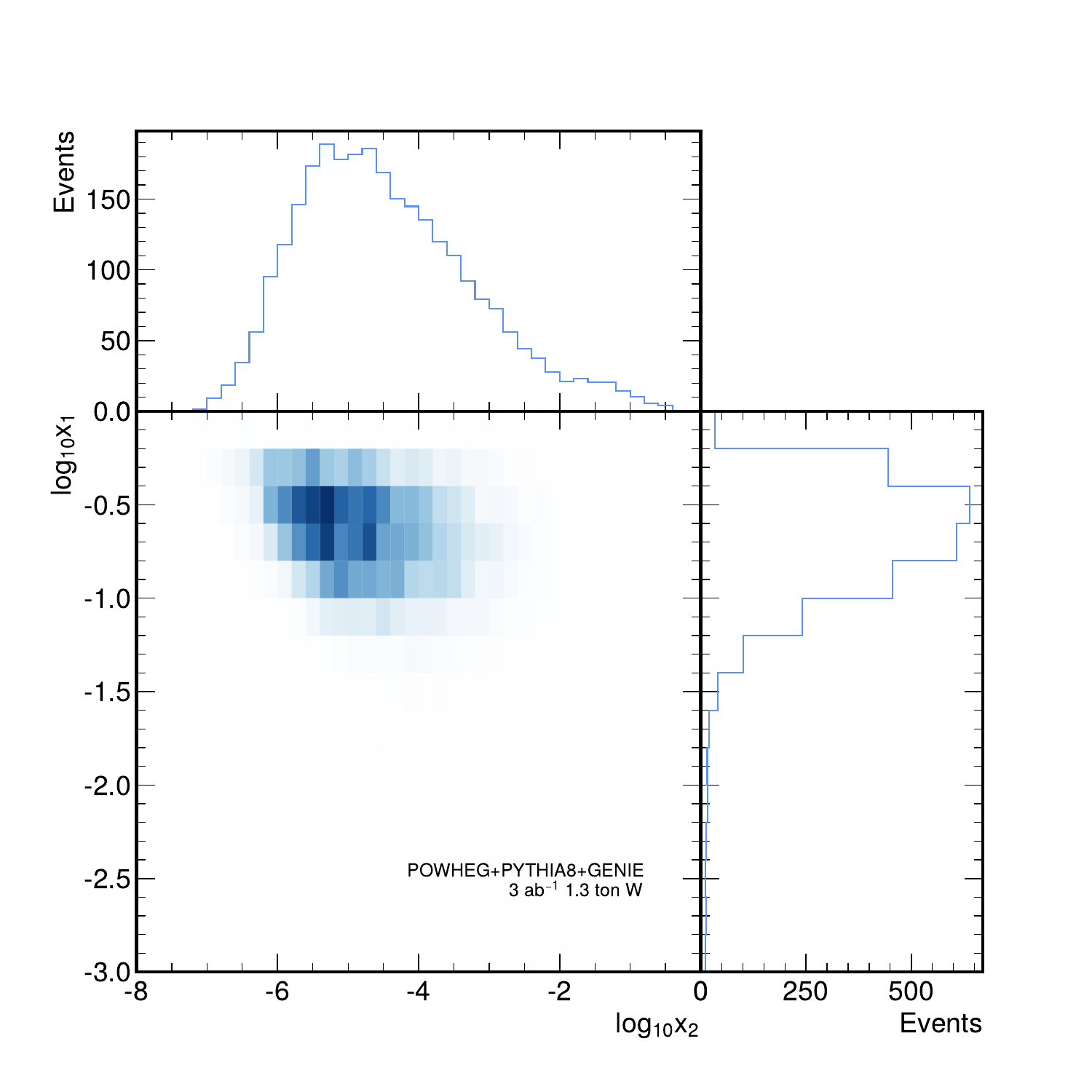}
	\caption{Distribution of Bjorken x of the colliding partons in open charm production events resulting in an  $\nu_e$ charged current interaction in SND@HL-LHC. The events were generated with \textsc{POWHEG}, with parton shower modeling by \textsc{Pythia8} and the  $\nu_e$ charged current cross section from \textsc{GENIE}.}
	\label{fig:nue_x}
\end{figure}

\begin{figure}[h!]
    \centering
	\includegraphics[trim={20 20 20 20}, clip, width=0.49\textwidth
	]{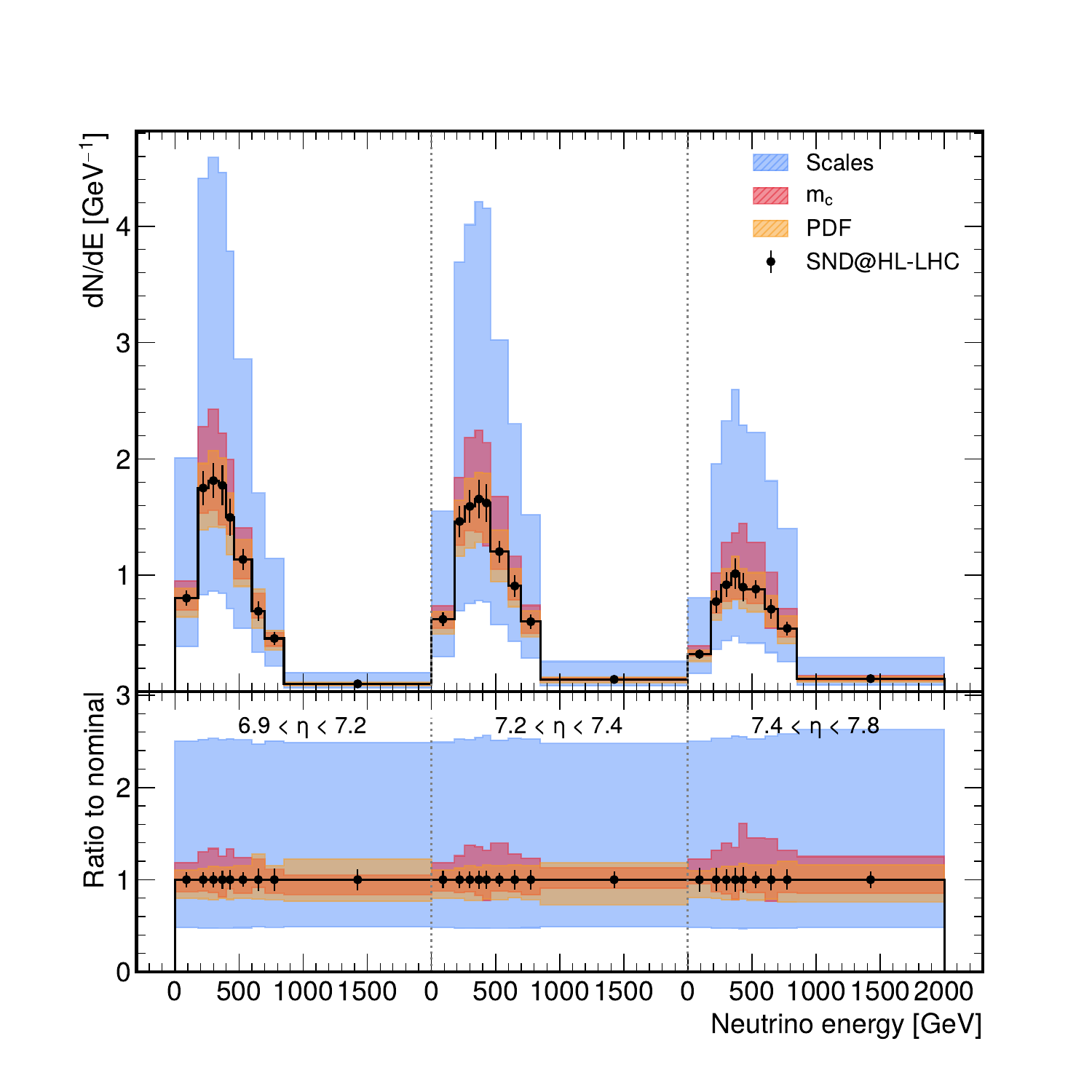}
	\includegraphics[trim={20 20 20 20}, clip, width=0.49\textwidth]{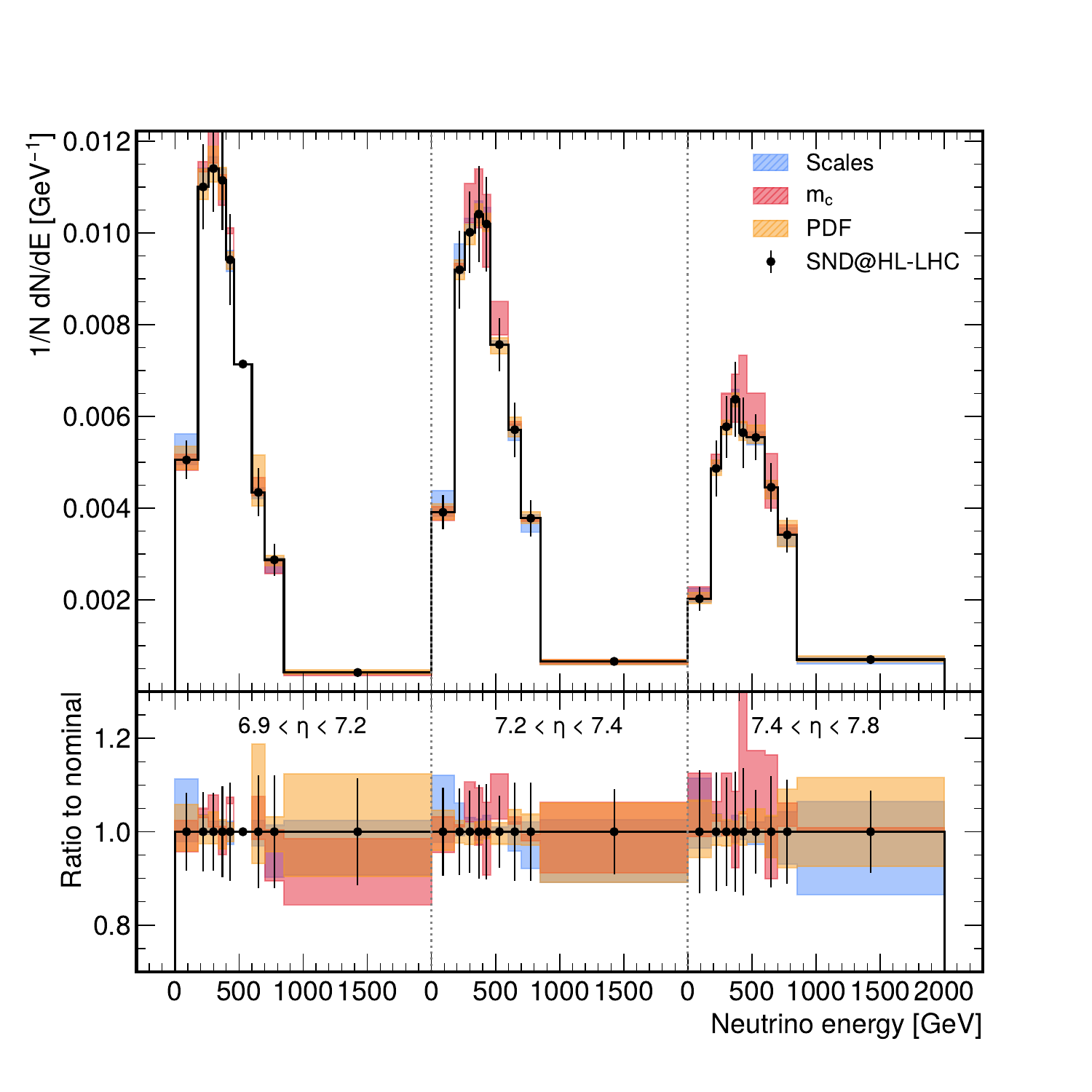}
	\caption{Distribution of energy and pseudo-rapidity of charged-current interactions of  $\nu_e$s of charm-hadron origin, simulated at parton level with \textsc{POWHEG}, with parton shower modeling by \textsc{Pythia8} and neutrino interaction cross sections from \textsc{GENIE}. Three sets of systematic uncertainty related to forward charm production are shown: the renormalization and factorization scales (varied to one half and double of the nominal values), the charm quark mass (varied between 1.25 and 1.65 GeV/c$^{2}$), and variations of the PDF based on Ref.~\cite{Buonocore:2023kna}. The figure on the left shows the event yield for 3000 fb$^{-1}$, and the figure on the right shows distribution normalized to an arbitrarily chosen reference bin \mbox{(460 $<$ E$_\nu$ $<$ 600 GeV, 6.9 $< \eta <$ 7.2)}, emphasizing the impact of systematic variations on the shape of the distribution.}
	\label{fig:gPDF}
\end{figure}

Electron neutrinos in the SND@HL-LHC pseudo-rapidity range $6.9<\eta<7.7$ are mostly produced by charm decays. Therefore, $\nu_e$s can be used as a probe of charm production in an angular range where the charm yield has a large uncertainty, to a large extent coming from the gluon parton distribution function (PDF). Electron neutrino measurements can thus constrain the uncertainty on the gluon PDF in the very small (below $10^{-5}$) Bjorken $x$ region (see Figure \ref{fig:nue_x}). The interest therein is two-fold: firstly, the gluon PDF in this $x$ domain will be relevant for Future Circular Collider (FCC) experiments; secondly, the measurement will reduce the uncertainty on the flux of very high energy ($\sim$PeV) atmospheric neutrinos produced in charm decays, essential for the evidence of neutrinos from astrophysical sources~\cite{Bai:2022xad},~\cite{Enberg:2016}.
The charm measurement by the current detector in Run\,3 will be affected by a systematic uncertainty of 35\% and by a statistical uncertainty of almost 10\%. The operation in HL-LHC of the SND@LHC detector will reduce the statistical uncertainty to about 2\%, as  is clear from Table~\ref{tab:nu_flux}. The combination of a large number of  $\nu_e$ events with charm origin, and the wide range of pseudo-rapidity covered by SND@HL-LHC allows for constraining PDFs using correlations of neutrino event rates between energy and pseudo-rapidity bins. Theory uncertainties that have a large impact on the overall event rate normalization, such as the renormalization and factorization scales and the charm quark mass, largely cancel out in shape-based measurements of the PDF, as demonstrated in Figure~\ref{fig:gPDF}. 

We estimate the systematic uncertainty of approximately 5\% in the gluon PDF measurement by considering the scale of non-PDF theory parameter variations within bins of the shape-only distribution.

The SND@HL-LHC detector’s ability to identify all three neutrino flavours enables testing Lepton Flavour Universality in neutrino interactions.

\begin{figure}[htbp]
\centering
\includegraphics[width=1.0\columnwidth]
{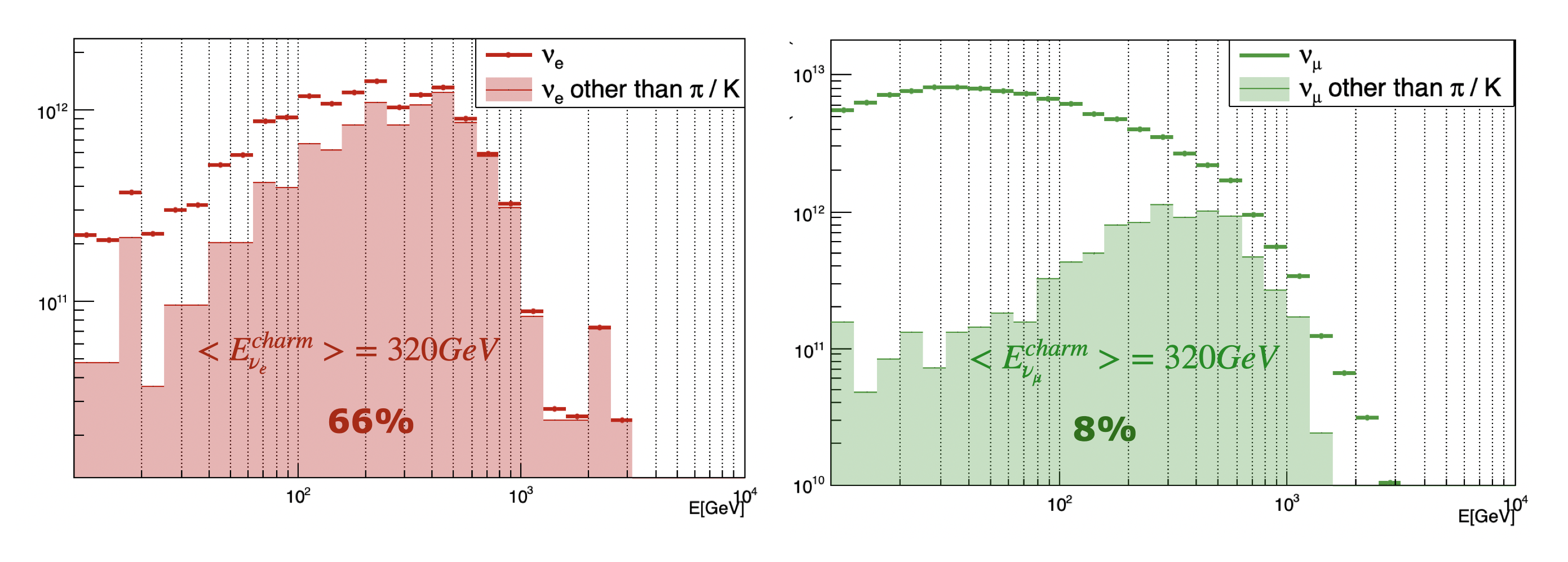}
\vspace{-1cm}
\caption{Energy spectrum of  $\nu_\mu$ (right) and  $\nu_e$ (left) neutrinos and anti-neutrinos in
the SND@LH-LHC acceptance. Filled areas represent the component coming from charm decays.}
\label{fig:nu_charm}
\end{figure}

Figure~\ref{fig:nu_charm} shows  $\nu_\mu$ and  $\nu_e$ spectra within the detector’s acceptance, with heavy-quark decay contributions as the filled area.

At the Target,  $\nu_e$s from pion and kaon decays constitute 34\% of the total flux but contribute only 20\% of interactions due to lower energies and cross-sections. Assuming tau and  $\nu_e$s originate from charmed hadron decays, the $\nu_e$ to $\nu_\tau$ ratio ($R_{13}$) depends on branching ratios and charm fractions, making it sensitive to their cross-section ratio and allowing an Lepton Flavour Universality test~\cite{SNDTP}.

Currently, a 30\% statistical uncertainty from low $\nu_\tau$ statistics dominates. At the HL-LHC SND@HL-LHC will reduce this to 6\% (Table~\ref{tab:nu_flux}). The main systematic uncertainty, 20\% from $D_s$ fragmentation functions, will improve with ongoing NA65~\cite{dstau}, SHiP-charm~\cite{Akmete:2286844}, and LHCb analyses.

Lepton flavour universality can also be tested using the  $\nu_e$-to- $\nu_\mu$ ratio ($R_{12}$). With an energy cut, charm can serve as a  $\nu_\mu$ source. Assuming a 600 GeV threshold in the SND@LHC pseudo-rapidity region, a 10\% accuracy is expected for both systematic and statistical uncertainties.

The $R_{12}$ measurement at SND@HL-LHC will be optimised to minimize contamination from light hadron decays while reducing statistical uncertainty to a few percent. The strategy involves constraining low-energy  $\nu_\mu$ flux via LHCf~\cite{LHCf:2015rcj}, achieving better than 10\% precision. The ratio will be measured in an energy and pseudo-rapidity region where charm dominates, optimizing the fiducial region to limit pion-decay contamination and reduce systematic uncertainty to a few percent.

The high neutrino beam energy results in most interactions occurring via Deep Inelastic Scattering (DIS).
DIS cross-sections have been measured at low energies ($E_\nu<350$ GeV) by beam dump experiments~\cite{ParticleDataGroup:2020ssz} and at high energies ($E_\nu>6.3$ TeV) by IceCube for  $\nu_\mu$s~\cite{IceCube:2017roe}.

SND@HL-LHC can measure cross-sections in the TeV range using  $\nu_\mu$s, with precise flux estimates from LHCf light meson production data~\cite{Adriani:2012ap}.
The magnetic spectrometer enables charge identification of muons from neutrino CC DIS interactions, allowing separate neutrino and anti-neutrino cross-section measurements up to 1 TeV. At higher energies, the average neutrino energy will be determined.

\begin{figure}[htbp]
\centering
\includegraphics[width=0.6\columnwidth]
{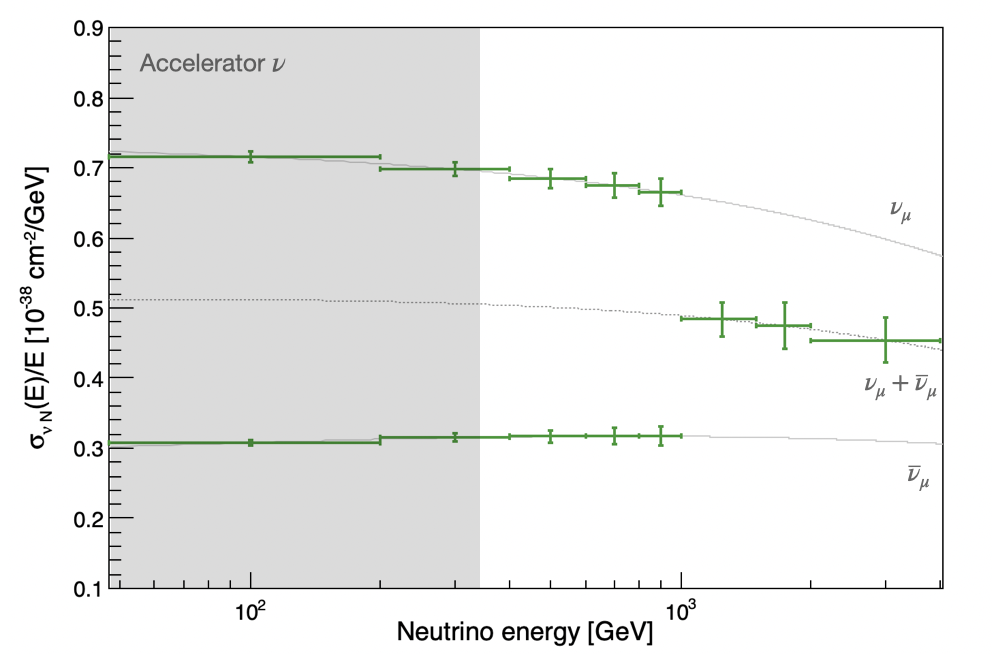}
\caption{Expected statistical uncertainties for  $\nu_\mu$ CC~DIS cross-section measurements. For energies below 1 TeV SND@HL-LHC can distinguish between muon neutrinos and anti-neutrinos. Theoretical predictions are evaluated using \textsc{GENIE} neutrino CC~DIS  cross-sections on a tungsten Target.}
\label{fig:nu_cross}
\end{figure}

Figure~\ref{fig:nu_cross} illustrates SND@HL-LHC’s capability to measure the $\nu_{\mu}N$ cross-section, with statistical uncertainties linked to Table~\ref{tab:nu_flux} neutrino yields.

The magnetic moment of neutrinos can be enhanced in some BSM models. The one of the $\nu_\tau$ is the less constrained due to the lack of data~\cite{DONUT:2001zvi}.  

A larger $\nu_\tau$ magnetic moment would lead to an increase in the cross section $\sigma_{\nu e}$ for the elastic scattering of neutrinos on electrons, which can be precisely calculated in the SM. An enhancement of $\sigma_{\nu e}$ would be a clear signal of new physics.
The signature of such events is the presence of a single electron in the final state, with peculiar kinematical features. The SND@HL-LHC detector can efficiently identify electromagnetic showers, and reconstruct their energy and direction.
Moreover, the $\nu_\tau$ flux can be estimated based on the measured yield of $\nu_\tau$ CC events, and the expected number of single-electron events according to the SM can be constrained by measurements of all flavour neutrinos CC and NC interactions on tungsten.
Given the large $\nu_\tau$ flux produced in p-p collisions at the HL-LHC, the SND@HL-LHC measurements can potentially constrain the $\nu_\tau$ magnetic moment.

\subsection{Charm-tagged neutrino events}

\begin{figure}
\centering
    \includegraphics[trim={10cm 8cm 6cm 1.2cm}, clip, width=0.6\textwidth]{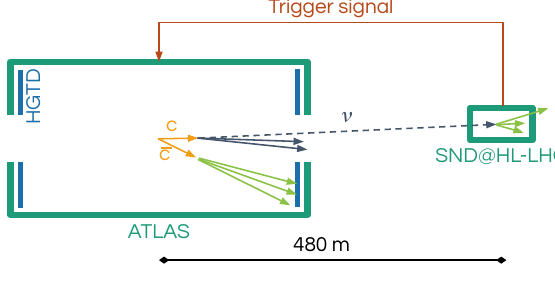}
    \vspace{-1cm}
    \caption{Sketch of a charm-tagged neutrino event in SND@HL-LHC and ATLAS.}
    \label{fig:charmtag_sketch}
\end{figure}

The high yield of neutrino events in SND@LHC enables the identification of a sub-sample coinciding with charm hadrons detected in ATLAS. While the neutrino's parent charm hadron is outside ATLAS acceptance, about 10\% of associated charm hadrons fall within it, allowing potential detection as sketched in Figure~\ref{fig:charmtag_sketch}. This charm-tagged neutrino sample provides clean flavor ratios with minimal light hadron contamination, enabling LFU tests with lower systematic uncertainty at the expense of  higher statistical uncertainty.

\textsc{Pythia8} simulations predict that each bunch crossing with 200 $pp$ collisions at HL-LHC, will produce four charm hadrons within ATLAS acceptance, spread over ~200 ps. Neutrino-charm matching requires timing resolution below 50 ps, achievable with ATLAS' Phase-II HGTD upgrade ($2.4 < \left| \eta \right| < 4.0$). SND@LHC must therefore have equally or more precise timing detectors.

\begin{figure}[hbtp]
	\includegraphics[trim={20 20 20 20}, clip, width=0.45\textwidth
	]{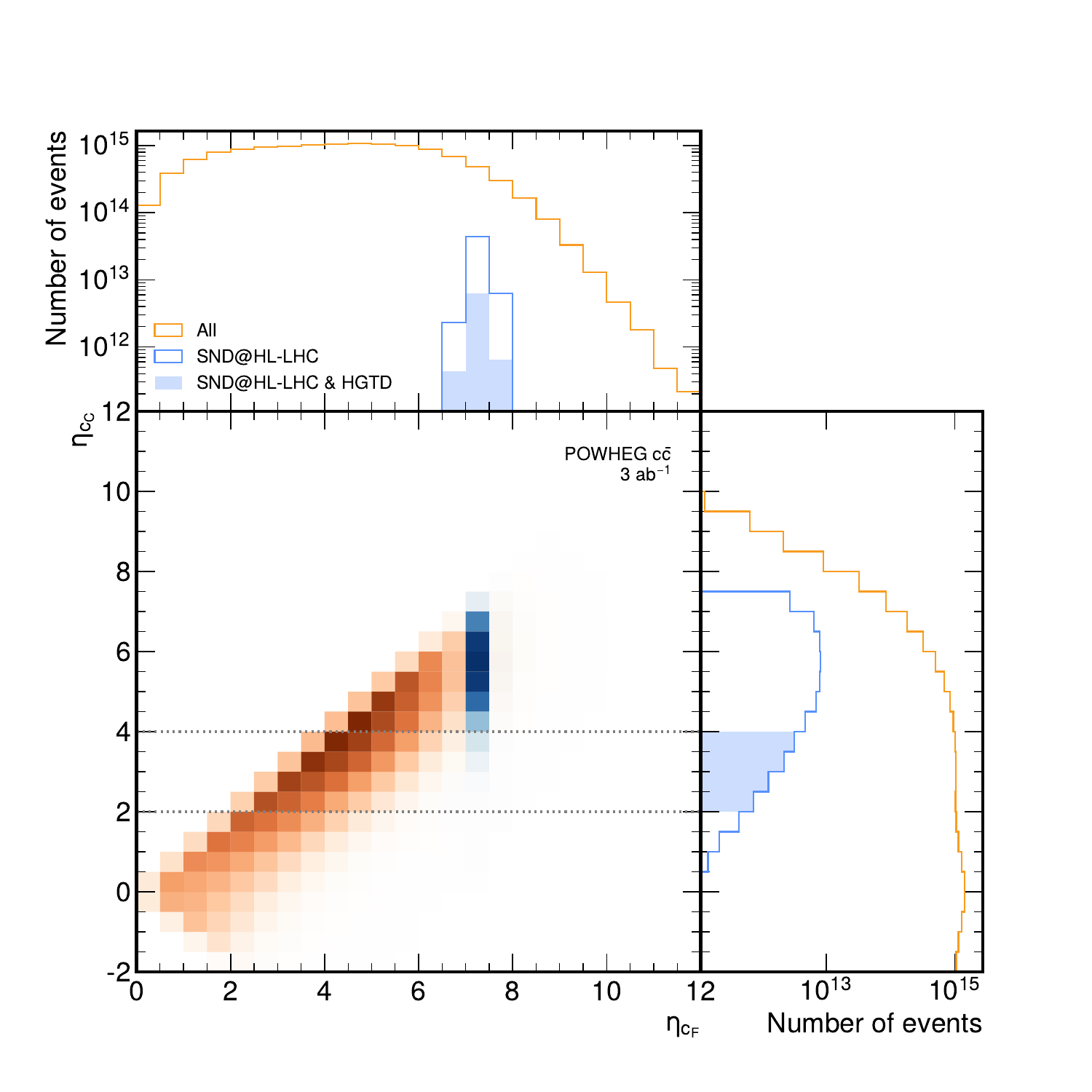}
	\includegraphics[trim={20 20 20 20}, clip, width=0.45\textwidth]{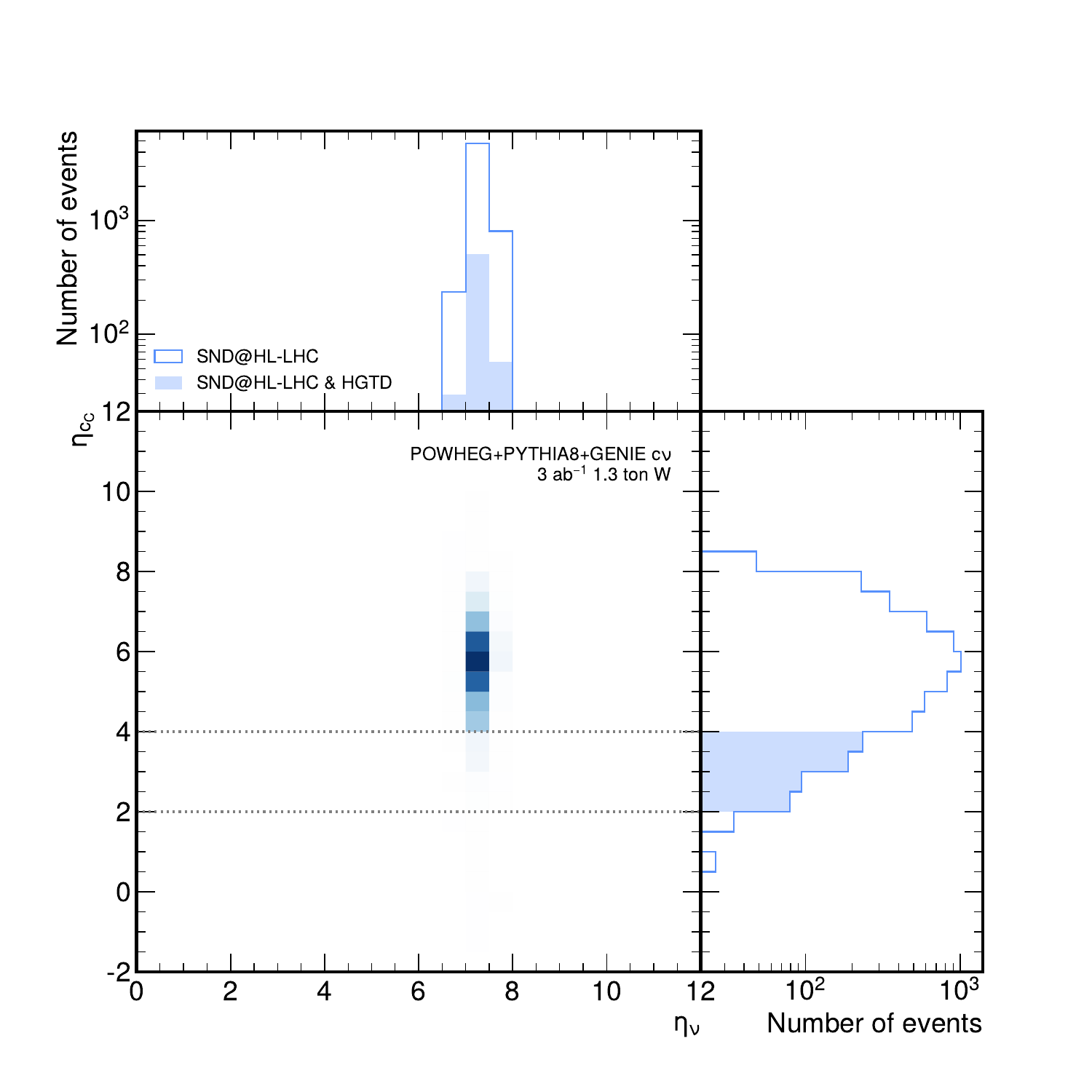}
	
    \caption{The left figure shows the pseudo-rapidity distribution of the most forward ($\eta_{c_F}$) and least forward ($\eta_{c_C}$) charm quarks in pairs from \textsc{POWHEG}. The right figure, including parton shower and neutrino interaction simulations from \textsc{Pythia8} and \textsc{GENIE}, displays the pseudo-rapidity of the most energetic neutrino ($\eta_{\nu}$) and $\eta_{c_C}$. The orange distribution represents the inclusive sample, while blue highlights events within the SND@HL-LHC acceptance. Shaded areas indicate events also detected by ATLAS HGTD. All distributions are normalized to 3 ab$^{-1}$ with a 1.3-ton neutrino target mass.}
	\label{fig:coincidentrate}
\end{figure}

The left panel of Figure~\ref{fig:coincidentrate} shows the pseudo-rapidity distribution of parton-level charm quark pairs from \textsc{POWHEG}. It highlights events where the most forward charm quark falls within SND@HL-LHC acceptance and the subset where the least central charm quark is within ATLAS HGTD acceptance. Of the forward charm events, 14\% also have the other charm quark in HGTD. The right panel includes \textsc{Pythia8} parton showers and \textsc{GENIE} neutrino interactions, showing that 11\% of interacting neutrino events contain a charm quark in HGTD acceptance. With 3000~fb$^{-1}$, 597 neutrino interactions in SND@LHC are expected with an associated charm hadron in ATLAS: 227 ($\nu_e+\bar{\nu}e$) CC, 212 ($\nu\mu+\bar{\nu}\mu$) CC, 10 ($\nu\tau+\bar{\nu}\tau$) CC, and 147 NC. These rates enable LFU testing via the charm-tagged ($\nu_e+\bar{\nu}e$)/($\nu\mu+\bar{\nu}\mu$) ratio with ~10\% statistical precision.

To ensure ATLAS retains data from particles produced alongside neutrinos interacting in SND@LHC, a trigger signal is sent from SND@LHC to ATLAS within the 10 µs latency of the upgraded ATLAS trigger. Given the 480 m round-trip time of ~3.2 µs, this signal arrives on time.

Extrapolating from Run 3 data, the expected neutral particle trigger rate at HL-LHC is ~1 Hz, negligible for ATLAS. Additionally, reducing light-jet background relies on machine learning-based charm-jet identification, which currently achieves nearly two orders of magnitude light-jet rejection with ~40\% charm-jet efficiency~\cite{Mondal:2024nsa,ATLAS:2022qxm,ATLAS:2022rkn}.

\vspace{-1cm}
\subsection{Summary of physics results with neutrinos}
\begin{table}
\centering
\begin{tabular}{l  c c |  c c }
\toprule
Measurement  &   \multicolumn{2}{c|}{Uncertainty}  & \multicolumn{2}{c}{Uncertainty}  \\
  &    Stat. & Sys. &   Stat. & Sys.\\
\midrule
Gluon PDF  & 5\% & 35\% & 2\% & 5\% \\
$\nu_e/\nu_\tau$ ratio for LFU test    & 30\% & 22\% & 6\% & 10\% \\
$\nu_e/\nu_\mu$ ratio for LFU test     & 10\% & 10\% & 2\% & 5\%\\
Charm-tagged $\nu_e/\nu_\mu$ ratio for LFU test     &- & - & 10\% & $<$ 5\%\\
$\nu_\mu$ and $\overline{\nu}_\mu$ cross-section & - & - & 1\% & 5\%  \\
\bottomrule
 \end{tabular}
 \caption{
 SND@HL-LHC neutrino interaction measurements with HL-LHC data vs. Run 3 estimates, including statistical and systematic uncertainties.
 }
  \label{tab:neutrino_physics}
 \end{table}
Table~\ref{tab:neutrino_physics} summarises the main HL-LHC physics objectives with the SND@HL-LHC detector in the analyses of neutrino interactions,  compared with estimates for the current detector in Run\,3. The proposed measurements are reported together with the estimated uncertainties, as described in detail in the corresponding sections. 

\section{Conclusion}
\begin{itemize}
    \item  
SND@HL-LHC will offer a significant contribution to the neutrino physics program at the LHC. 
     \item 
Upgrading the detector requires a modest investment ($<1$ MCHF) by the Collaboration. Parts of the original CMS tracker will be repurposed. 
\item
SND@HL-LHC will test neutrino interactions at the highest accelerator energies, offer unprecedented probes of lepton flavor violation in the neutrino sector, and enhance sensitivity to Beyond Standard Model effects. Additionally, it will improve our understanding of previously unexplored QCD domains and provide valuable insights into charm quark production and the intricacies of its hadronization.
\end{itemize}

A recognition and endorsement of the SND@HL-LHC programme  will be highly beneficial to its successful implementation.

\newpage 
\newpage 

\clearpage
\bibliographystyle{JHEP} %
\bibliography{references}

\end{document}